\documentclass{PoS}

\usepackage{verbatim}

\newcommand{\pHe}{\textsf{p/He}}

\newcommand{\He}{\textsf{He}}
\newcommand{\Li}{\textsf{Li}}
\newcommand{\Be}{\textsf{Be}}
\newcommand{\B}{\textsf{B}}
\newcommand{\C}{\textsf{C}}
\newcommand{\N}{\textsf{N}}
\newcommand{\Oxy}{\textsf{O}}
\newcommand{\Fe}{\textsf{Fe}}
\newcommand{\BC}{\textsf{B}/\textsf{C}}
\newcommand{\eplus}{\ensuremath{e^{+}}}
\newcommand{\pfrac}{e\ensuremath{^{+}}/(e\ensuremath{^{-}}\,+\,e\ensuremath{^{+}})} 
\newcommand{\CO}{\textsf{C/O}}
\newcommand{\OFe}{\textsf{O/Fe}}
\newcommand{\CFe}{\textsf{C/Fe}}
\newcommand{\NeFe}{\textsf{Ne/Fe}}
\newcommand{\MgFe}{\textsf{Mg/Fe}}
\newcommand{\ArFe}{\textsf{Ar/Fe}}
\newcommand{\CaFe}{\textsf{Ca/Fe}}

\newcommand{\ApJ}{ApJ}
\newcommand{\AeA}{A\&A}
\newcommand{\PRL}{PRL}
\newcommand{\PRD}{PRD}
\newcommand{\etal}{et alii}

\newcommand{\AMS}{\textsf{AMS}}
 
\newcommand{\ie}{\textit{i.e.}}

\def\Journal#1#2#3#4{{#4}, {#1}, {#2}, #3}

\title{Consistent description of leptonic and hadroninc spectra in cosmic rays}

\ShortTitle{Leptons, Hardons and Nuclei in Cosmic Rays}

\author{\speaker{Nicola Tomassetti}\thanks{A footnote may follow.}\\
LPSC, Universit\'e Grenoble-Alpes, CNRS/IN2P3, F-38026 Grenoble, France; email: nicola.tomassetti@lpsc.in2p3.ch\\
        E-mail: \email{nicola.tomassetti@lpsc.in2p3.fr}}

\abstract{
The \AMS{} Collaboration has recently released data on cosmic ray (CR) leptons and hadrons that can shed light 
on two exciting problems in CR physics: on one side, the origin of the rise of the CR positron fraction 
above 10 GeV of energy, on the other side, the nature of the spectral features observed in CR protons and 
helium at ~TeV energies. 
Concerning heavier nuclei, 
The ATIC-2 experiment has recently reported an puzzling spectral upturn at energy $\gtrsim$\,50\,GeV per nucleon
in several primary/primary ratios involving Iron, such as the \OFe{} or \CFe{} ratio.
In this work, the AMS data are described using a two-component scenario, where the 
total CR flux is provided by a mixture of fluxes accelerated by sources with different properties. 
Within this picture, the role of secondary CR production inside nearby supernova remnants is discussed.
In particular, we present the predictions of our model for the \CFe{} and \OFe{} ratios, in 
connection with the spectral anomalies found by the ATIC-2 experiment.
}

\FullConference{The 34th International Cosmic Ray Conference,\\
		30 July- 6 August, 2015\\
		The Hague, The Netherlands}

\begin{document}

\section{Introduction}

The observed cosmic ray (CR) spectrum has several puzzling features such as
the 10--200\,GeV rise of the positron fraction \pfrac{} \cite{Adriani2009,Serpico2011},
the spectral hardening of proton and helium at energy $\gtrsim$\,300\,GeV/nucleon \cite{Adriani2011,Yoon2010,Blasi2013},
or the unexpected decrease of the \pHe{} ratio as function of rigidity, $R=p/Z$, between $R\approx$\,10\,GV and $R\approx1$\,TV.
Recently, a puzzling spectral upturn has been reported by the ATIC-2 experiment for nuclear ratios involving
Iron, such as the \CFe{} or \OFe{} ratios at $\sim$\,50\,GeV/nucleon of energy \cite{Panov2013,Panov2014}. 
Many of these features are now being investigated with high precision by the \AMS{} experiment \cite{Aguilar2015,Accardo2014}.

In the standard model descriptions, \emph{primary} CRs such as electrons, protons, \He, \C-\N-\Oxy, or \Fe{} nuclei 
are released in the interstellar medium (ISM) by a continuous distribution of supernova remnants
after being accelerated via diffusive-shock-acceleration to power-law spectra $\sim\,E^{-\nu}$, with $\nu\approx$\,2--2.4, up to PeV energies.
The spectrum injected in the ISM is therefore steepened by diffusive propagation in the halo (typical half-size $L\sim$\,3-10\,kpc)
where the diffusion coefficient is assumed to be $K\propto E^{\delta}$, with $\delta\sim$\,0.3--0.7.
Collisions of CRs with this ISM nuclei of the Galactic disk (half-size $h\sim$\,100\,pc) give rise 
to \emph{secondary} species such as \eplus{} or \Li-\Be-\B{} elements, that are expected to be
$E^{\delta}$-times steeper than primary CRs.
The several models based on this understanding agree in predicting power-law spectra for primary nuclei, 
smooth primary/primary ratios, and a decreasing positron fraction \cite{Strong2007}. 
The recently observed features in CR protons, nuclei, and leptons are at tension with these predictions.
The positron excess requires ad additional $e^{\pm}$ component that may be emitted by nearby exotic sources, 
such as dark-matter annihilation/decay, or by known astrophysical sources, such as pulsars or \emph{old} 
supernova remnants (SNRs) \cite{Serpico2011}.  

In the \emph{old SNR} scenario, the positron excess is produced by collisions of accelerated protons
in proximity of the shock \cite{Blasi2009}. The $e^{\pm}$ production and their subsequent acceleration 
gives rise to a SNR component which is \emph{one power harder} than that of primary protons or electrons, 
$E^{-\nu}$, and may potentially explain the positron fraction anomaly \cite{MertschSarkar2014}.
From the same mechanism, other secondary nuclei such as \Li-\Be-\B{} or antiprotons are expected to be produced.
if that the observed CR flux is entirely made up by this nearby SNR, 
one has to expect \emph{rise} of the \BC{} ratio at $\sim$\,50\,GeV per nucleon \cite{MertschSarkar2009}.
However, recent measurements of the \BC{} ratio do not show any rise \cite{CholisHooper2014}.

\section{A two-component scenario with nearby source}

In \cite{TomassettiDonato2015}, we have argued that the \emph{old} SNR scenario is incomplete 
to account for the CR data on hadronic spectra at multi-TeV energies.
Acceleration at TeV-PeV energies can be only achieved with a magnetic field amplification mechanism which,
however, is very hardly compatible with secondary production at the shock \cite{Serpico2011,Kachelriess2011}. 

Furthermore, the spectral hardening of CR proton and helium may suggest the emergence
of a different source on their high-energy flux \cite{Vladimirov2012}.
We have therefore proposed a two-component scenario where the total CR flux is 
provided by a \emph{local} source component $\phi^{\rm L}$ in the $\sim$\,GeV--TeV region, 
arising from a \emph{old} SNR, and by a \emph{Galactic ensemble} SNR component $\phi^{\rm G}$, 
emitted by younger SNRs with amplified magnetic fields, in the $\sim$\,TeV--PeV region. 
An important consideration is that, due radiative losses, the $e^{\pm}$ propagation length 
is limited to a typical distance $\lambda^{\rm rad} \sim \sqrt{\tau^{\rm rad} K} \propto E^{(\delta-1)/2}$
with characteristic time $\tau^{\rm rad}\sim 300\cdot E^{-1}$\,Myr\,GeV$^{-1}$.
A local source placed within a few 100\,pc is therefore necessary to explain the GeV-TeV $e^{\pm}$ flux \cite{Delahaye2010}.
In contrast, protons and nuclei do not experience radiative losses. Their local flux may 
arise from a larger population of Galactic sources \cite{Taillet2003}.
As we have shown, a two component scenario may account for the rise in the positron fraction 
\emph{and} for the decreasing of the \BC{} ratio. 

\begin{figure}[!t] 
  \includegraphics[width=0.45\textwidth]{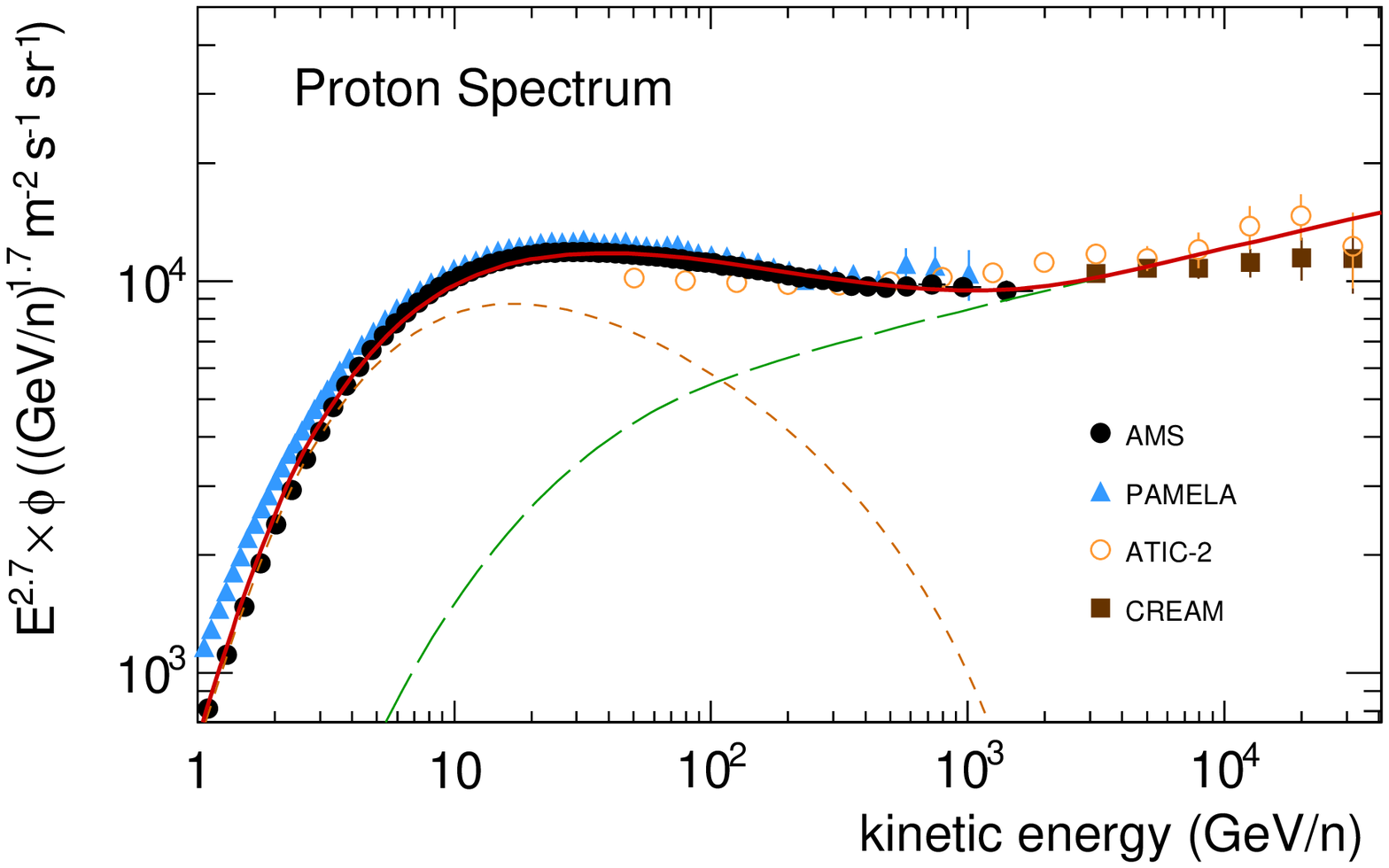}
  \qquad
  \includegraphics[width=0.45\textwidth]{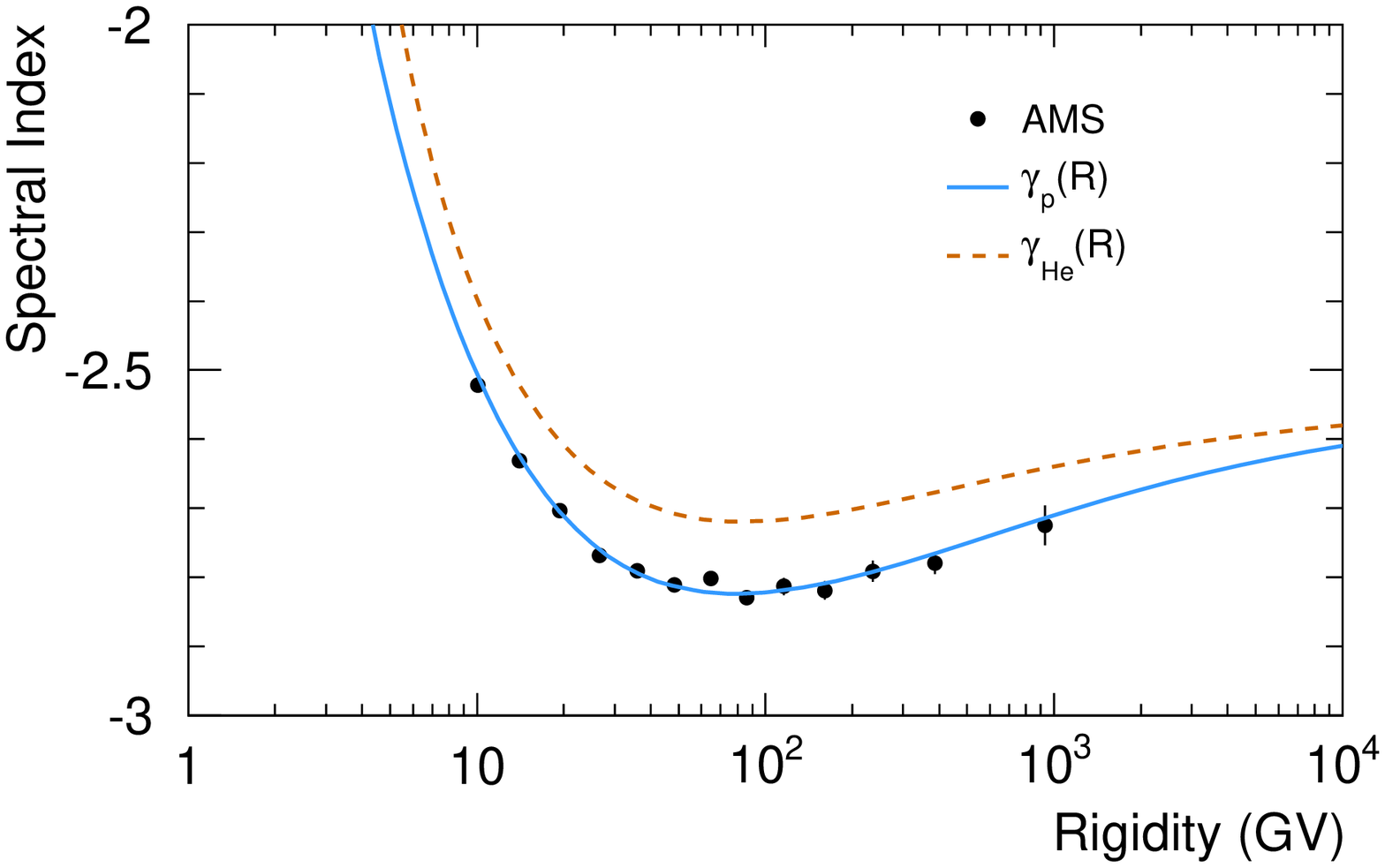}
\caption{\footnotesize%
  Left: Energy spectrum of CR protons multiplied by $E^{2.7}$.
  The solid lines indicate the model calculations. The contribution arising from
  the nearby SNR (short-dashed lines) and from the Galactic SNR ensemble (long-dashed lines) are shown.
  The data are from \AMS{} \cite{Aguilar2015}, PAMELA \cite{Adriani2011}, ATIC-2 \cite{Panov2009}, and CREAM \cite{Yoon2010}. 
  Right: Proton an \He{} differential spectral index from the model in comparison with the \AMS{} proton data.
}\label{Fig::ccProtonSpectrum} \label{Fig::ccPositronFlux}
\end{figure}

In Fig.\,\ref{Fig::ccProtonSpectrum} we show our model with the new \AMS{} data on the proton spectrum.
These new data are well consistent with a smooth flux transition as described by model.
The two components of the flux are shown, \ie, split 
as $\phi_{H}=\phi_{H}^{\rm L} + \phi_{H}^{\rm G}$. Similarly for helium, one has $\phi_{He}=\phi_{He}^{\rm L} + \phi_{He}^{\rm G}$. 
The \AMS{} data show a differential spectral index which progressively 
hardens at rigidity above $\sim$\,100\,GV \cite{Aguilar2015}.
It is also interesting to compare the rigidity dependence of the differential
spectral index $\gamma(R)=d[\log(\phi)]/d[log(R)]$. The functions $\gamma_{p}$ and $\gamma_{He}$ from the 
model are shown in Fig\,\ref{Fig::ccProtonSpectrum} in comparison with the \AMS{} proton data.
The model describes very well the smooth evolution of the spectral index at in the 10--1000\,GV rigidity range.
At higher rigidities, it can be seen that both species converge asymptotically to the same value, $\gamma^{\rm G}\cong -\nu^{\rm G}-\delta$=\,$-$2.6.


%
%
%

\section{B/C ratio}

As discussed, a two-component model gives a good description of the primary CR spectral hardening
which is now determined as the superposition of $\phi_{H}^{\rm L}$ and $\phi_{H}^{\rm G}$.
Similar superposition is expected for the fluxes of primary nuclei such as the carbon flux,which
will be also of the type $\phi_{\rm C}\approx \phi_{\rm C}^{\rm L} + \phi_{\rm C}^{\rm G}$.
Thus, it also experiences a spectral hardening that is well reproduced by the model. 
The \B{} spectrum, entirely secondary, can be ideally split into $\phi_{\rm B}=\phi_{\rm B}^{\rm L} + \phi_{\rm B}^{\rm ISM}$,
where the first component is the one produced inside the OSNR, and the second arises in the ISM via 
collisions of heavier nuclei such as \C, \Oxy, or \Fe. 
The ISM component $\phi_{\rm B}^{\rm ISM}$ can be split again into $\phi_{\rm B}^{\rm ISM/L}$ 
(produced by collisions of nuclei emitted by the local component) and $\phi_{\rm B}^{\rm ISM/G}$ 
(from collisions of primary nuclei injected by the Galactic ensemble).
In previous works of this kind, such as \cite{CholisHooper2014}, the \BC{} ratio is always meant 
as $(\phi_{\rm B}^{\rm L}+\phi_{\rm B}^{\rm ISM/L})/\phi_{\rm C}^{\rm L}$, because the total CR flux
is assumed to be provided only by one type of source.
In these works, the ratio $(\phi_{\rm B}^{\rm L}+\phi_{\rm B}^{\rm ISM/O})/\phi_{\rm C}^{\rm L}$ starts rising at $E \gtrsim$\,100\,GeV per nucleon
In our two-component scenario, the \BC{} ratio $\phi_{\rm B}/\phi_{\rm C}$ is found to decrease with energy in good agreement with the data. 
The trend of our \BC{} ratio is similar to that of  standard propagation models where only the Galactic component is considered.
In fact, at $\sim$\,100\,GeV/nucleon, when the local (harder) component of \B{} would become relevant enough to provide a signature
on the \BC{} ratio (\ie, $\phi_{\rm B}^{\rm L} \gtrsim \phi_{\rm B}^{\rm ISM/O}$), the total fluxes of \B{} and \C{} become
both dominated by the Galactic ensenble component $\phi_{\rm B}^{\rm ISM/G}$ and $\phi_{\rm C}^{\rm G}$.

\section{A look to heavier nuclei: C/Fe and O/Fe ratios}  

Here I show some important consequence of the two-component scenario for 
the prediction of heavy nuclei spectra, such as the Iron energy spectrum.
Under a scenario where the flux is a superposition between a local source component and the
flux of more distant sources, the shape of the spectral transition between the two compontents
has a characteristic signature in the spectrum of heavy nuclei, due to a combination of propagation and spallation effects.
In fact, the propagation range of heavy nuclei like Iron is spatially limited 
by inelastic collisions with the ISM nuclei, which may prevent CRs injected from distant sources to reach the Solar System.
To study this effect, I make use of an effective calculation scheme, based on the propagation scale length, 
that enables a prediction for the ratios between light and heavy primary nuclei such as \CFe{} and \OFe. 
For these ratios, a remarkable spectral upturn is predicted at $\sim$\,50\,GeV of energy.
\begin{figure}[!t] 
  \includegraphics[width=0.80\textwidth]{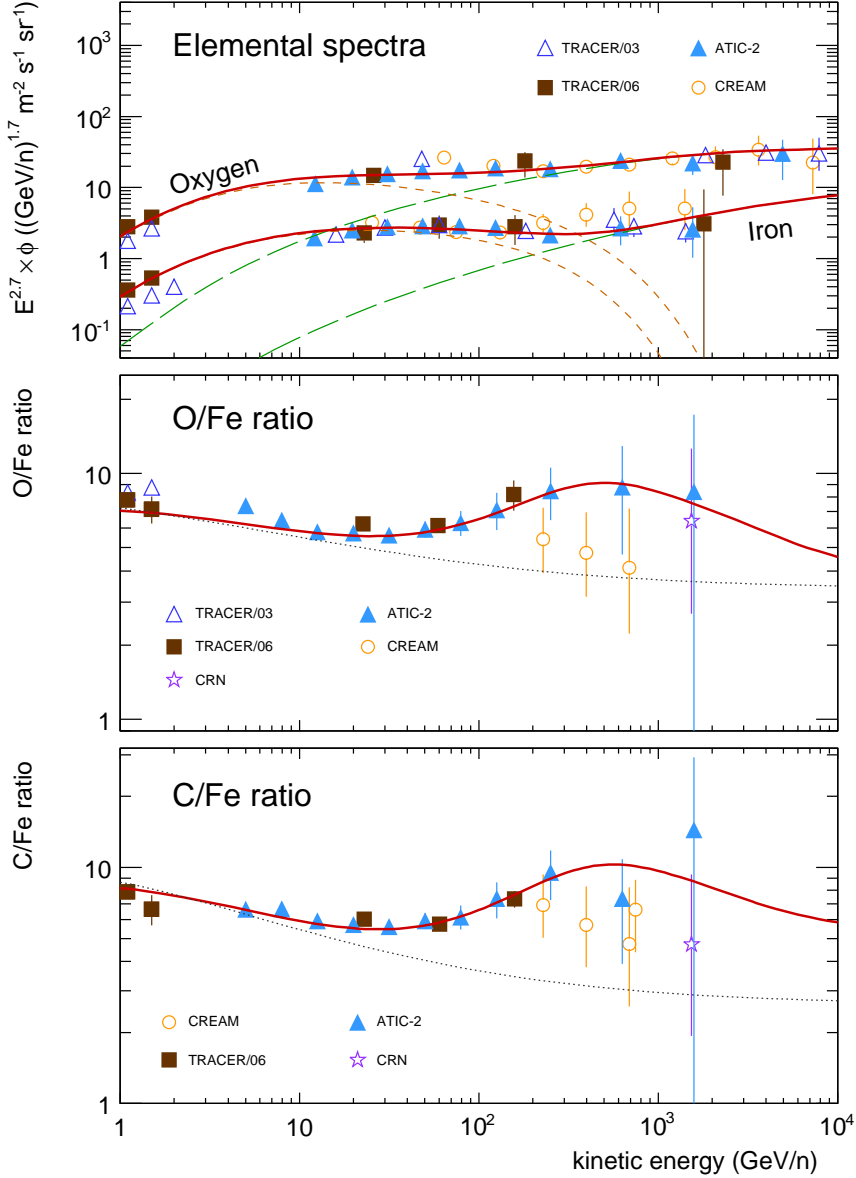}
\caption{\footnotesize%
  Left: distribution function (solid line) and cumulative function (dashed line) of Galactic SNRs as function of the distance $d$ from the solar system, and fraction of Galactic SNRs contributing to the CR flux of \Oxy{} (solid line) and \Fe{} (dashed line).
  Right: Energy spectra of \Oxy{} and \Fe{} multiplied by $E^{2.7}$, and nuclear ratios
  \CFe{} and \OFe{} as function of kinetic energy per nucleon.
  The solid lines indicate the model calculations. The contributions arising from the
  nearby source (short-dashed lines) and from the Galactic ensemble (long-dashed lines) are shown.
  The data are from ATIC-2 \cite{Panov2013,Panov2014}, CREAM \cite{Ahn2009}, CRN \cite{Muller1991}, and TRACER \cite{Ave2008,Obermeier2011}. 
  The TRACER data on nuclear ratios are those obtained in Ref.\cite{Panov2014}.
  Standard model predictions are also shown for the \CFe{} and \OFe{} ratios (dotted lines).
}\label{Fig::ccHeavyNucleiFlux}
\end{figure}
%
%
In standard diffusion models, as long as all the contributiong CR sources have the same spectra,
the model predictions for the spectra of CR nuclei at Earth are
almost unsensitive to the exact distribution of Galactic sources.
Conversely, if the CR flux is produced by different classes of sources characterized by different spectra
and composition, it becomes very important to account for the source spatial distribution \cite{Taillet2003}.
This is the case for the propagation of heavy nuclei in our two-component scenario
with source components  $S^{\rm L}$ and $S^{\rm G}$, that accelerate CR particles to different spectral shapes.
In particular, the Galactic ensemble component $S^{\rm G}$ represents the contribution 
of a large-scale SNR population that extend to several kpc of distance.
Thus, one may argue that the fraction of Galactic SNRs effectively contributing to the 
local observed flux will depend on the propagation properties of the considered element.
Roughly, the typical propagation distance of CR nuclei, $\lambda^{\rm sp}\equiv\sqrt{K\tau^{\rm sp}}$, 
can be estimated from the spallation time-scale $\tau^{\rm sp}\cong \frac{L}{h\Gamma^{\rm sp}}$.
In this approximation the $h/L$ ratio accounts for the fact that CRs interact with the matter
only where they are in disk, \ie\, the total matter density is considered as averaged 
in the propagation halo \cite{Jones1978}.   
It interesting to note that, in contrast to CR leptons, the function $\lambda^{\rm sp}$ for 
nuclei \emph{increases} with energy and \emph{decreases} with mass. 
The mass dependence arise from the total interaction cross sections, $\sigma$, that 
depend on the projectile mass roughly as $\sigma^{\rm sp} \propto M^{0.7}$ \cite{Letaw1983}.
This gives $\lambda^{\rm sp}(E) \propto M^{-0.35}E^{\delta/2}$. 
This trend shows that, for CR spectrum of nuclei expected at Earth,
heavier species must come from sources located in nearer regions.
From these considerations, the fraction of Galactic sources that effectively contribute
to the observed flux at Earth have to be reduced for heavier nuclei. 
In order to estimate this fraction, one has to model a realistic spatial 
distribution for the Galactic SNRs, possibly as function of the distance from Earth.  
For this purpose, I follow closely the effective approach of \cite{Ahlers2009}.
Using the Monte-Carlo method and a realistic model of the Galactic structure as input,
I have computed the SNR probability density function $P(d)$ as function of the 
distance $d$ from the Earth.
This probability, which reflects the spatial distribution of the SNR Galactic ensemble.
is suppressed within $\sim$\,1.5\,kpc due to the inter-arm position of the Solar System.
The cumulative fraction of SNRs placed within a certain distance $d$ can be computed as $F(d)=\int_{0}^{d}P(l) dl$.
Thus, the fraction of Galactic SNRs contributing to the CR flux of a $j$-type element detected at Earth
is estimated as $F(\lambda^{\rm sp}_{j})$, where the distance $\lambda^{\rm sp}_{j}$ can be
eventually expressed as function rigidity or energy.
Thus, we replace source term of the Galactic SNR component, $S_{j}^{\rm G}(E)$, with the
\emph{effective source term} $\hat{S}_{j}^{\rm G}\equiv F_{j}(E)\times S_{j}^{\rm G}(E)$ for all $j-$types nuclei.
As discussed, $F_{j}(E)\equiv F(\lambda^{\rm sp}_{j}(E))$ is the fraction of Galactic SNRs contributing the nuclear species $j$-th.
At high energies one has $\hat{S}^{\rm G}\rightarrow S^{\rm G}$ for all particles, but the convergence is slower for heavier CR nuclei. 
Typical values of inelastic crosssections on hydrogen target are $\sigma^{\rm sp} \sim$\,40\,mb for protons, 
$\sim$\,300\,mb for \C-\N-\Oxy{}, and $\sim$\,900\,mb for \Fe.
The propagation setup of the model follows closely our earlier work \cite{TomassettiDonato2015,TomassettiDonato2012}. 
More details of this study will be presented in a forthcoming work \cite{Tomassetti2015tsnr}.
The model predictions for the \OFe{} and \CFe{} ratios and for the spectra of \Fe{} and \Oxy{}
are shown in Fig.\,\ref{Fig::ccHeavyNucleiFlux}. The two flux components are shown.
%
Due spallation, the \Fe{} spectrum of the Galactic SNR population is slightly re-shaped, because of
the ``missing flux'' from distant SNRs, \ie, SNRs that \emph{do not} contributed to the total \Fe{} flux at Earth.
This effect is maximized when plotting the primary/primary ratios \CFe{} and \OFe{} that. 
As seen in the figure, these ratios show a spectral upturn above a few tenths of GeV/nucleon energies. 
The ATIC-2 data are described very well by our model.
These spectral upturn is also present in the TRACER data, as shown in \cite{Panov2014}, but not in the CREAM data.
%
Future precision data will clarify the situation.
These features are almost canceled out in the \CO{} ratio, which does not show prominent features.
At low energy ($E\sim$\,1--10 GeV/nucleon), the flux is entirely dominated by the local source component. 
At high energy  ($E\gtrsim$\,TeV/nucleon) the spallation effect vanishes, so that the \CFe{} and \OFe{} 
ratios become representative of the spectral properties of the Galactic ensemble.
The \CFe{} and \OFe{} ratios predicted from the standard model are also shown in the figure (dotted line)
for comparison. Standard models make use of only one class of sources.
As discussed, the standard diffusion models are unable to describe the spectra upturn of the ATIC-2 data.
From this mechanism, we expect similar features in other ratios involving heavy nuclei, such as \NeFe, \MgFe, or \ArFe. 
It is interesting to note that an upturn in the \ArFe{} and \CaFe{} ratios was observed by the HEAO3 experiment 
but, at that time, it was not recognized as a real physical effect \cite{Binns1988}.
%
%

\section{Conclusions} 

This work is motivated by the search of a model for Galactic CRs that is able to account for the 
several puzzling observations in their spectrum.
Recent unexplained features of the CR spectrum are the rise in the positron fraction, 
the spectral change of slope in the CR hadrons, or the unexpected spectral upturn in the \CFe{} and \OFe{} ratio. 
Also interesting is the \pHe{} anomaly, \ie, the unexpected difference found between proton
and helium energy spectra. The possible connection of this anomaly with our scenario will be discussed in a future work.
We want to note that the presence of a nearby SNR, placed near the Solar System,
has been proposed in many CR physics studies \cite{Malkov2012,Vladimirov2012,Moskalenko2003}. 
Independent evidence for such a source have been found in studies of the Local Bubble \cite{Benitez2002,Moskalenko2003}. 
Though a nearby source may be not be detectable in $\gamma$--rays any longer, it can still contribute to the flux of CRs.
Without nearby sources, 
In the framework of the conventional models of CR propagation, it is difficult to interpret the ATIC-2 data.
The present study, however, relies in a number of simplified assumptions so that a more realistic 
treatment is needed. On the other hand, a rigorous treatment of the problem has to account for several unknown
that require the use of more precise data on CR nuclei, and in particular on the \Fe{} spectrum.
Luckily enough, this is a proficient era for CR physics. Along with \AMS, currently operating on the
International Space Station, the forthcoming space experiments such as
ISS-CREAM \cite{Seo2014}, or CALET \cite{Adriani2014Calet} will be able to provide 
high quality data on CR elements over a large energy range. 

{\footnotesize%
  This work is supported by the LabEx grant \textsf{ENIGMASS}. 
}


\begin{thebibliography}{99}


\bibitem{Adriani2009} Adriani,~O.,~\etal,
\Journal{Nature}{458}{607}{2009}
(PAMELA)

\bibitem{Serpico2011} Serpico,~P.,
\Journal{Astropart.Phys.}{39-40}{2--11}{2011}

\bibitem{Adriani2011} Adriani,~O.,~\etal,
\Journal{Science}{332}{6025}{2011}
(PAMELA)

\bibitem{Yoon2010} Yoon,~H.,~S.,~\etal,
\Journal{\ApJ}{715}{1400--1407}{2010}
(CREAM)

\bibitem{Blasi2013} Blasi,~P.,
\Journal{\AeA{} Rev.}{21}{70}{2013}

\bibitem{Accardo2014} Accardo,~L.,~\etal, 
\Journal{\PRL}{113}{121101}{2014}
(\AMS)

\bibitem{Aguilar2015} Aguilar,~M.,~\etal, 
\Journal{\PRL}{114}{171103}{2015}
(\AMS)

\bibitem{Panov2013} Panov,~A.~D.,~\etal,
J. Phys.: Conf. Series, 2013, 409, 012036
(ATIC-2)                        

\bibitem{Panov2014} Panov,~A.~D.,~\etal,
Nucl. Phys. B, 2014, 256-257, 223
(ATIC-2)                        

\bibitem{Strong2007} Strong,~A.~W., \etal, 
\Journal{Ann.Rev.Nucl.\&Part.Sci.}{57}{285--327}{2007}

\bibitem{Blasi2009} Blasi,~P.,
\Journal{\PRL}{103}{051104}{2009}

\bibitem{MertschSarkar2014} Mertsch,~P., \&~Sarkar,~S.,
\Journal{\PRD}{90}{061301}{2014}

\bibitem{MertschSarkar2009} Mertsch,~P., \&~Sarkar,~S.,
\Journal{\PRL}{103}{081104}{2009}

\bibitem{CholisHooper2014} Cholis,~I., \& Hooper,~D.,
\Journal{\PRD}{89}{043013}{2014} 

\bibitem{TomassettiDonato2015} Tomassetti,~N., \& Donato,~F.,
\Journal{\ApJ}{803}{L15}{2015} 

\bibitem{Kachelriess2011} Kachelrie\ss,~M., \etal, 
\Journal{\ApJ}{733}{119}{2011}

\bibitem{Vladimirov2012} Vladimirov, A. E., \etal,
\Journal{ApJ}{752}{68}{2012}

\bibitem{Delahaye2010} Delahaye,~T., \etal,
\Journal{AeA}{524}{A51}{2010}

\bibitem{Taillet2003} Taillet,~R. \& Maurin,~D., 
\Journal{\AeA}{402}{971--983}{2003} 

\bibitem{Panov2009} Panov,~A.~D.,~\etal,
\Journal{Bull.Russ.Acad.Sci.}{73-5}{564--567}{2009} 
(ATIC-2)                        

\bibitem{Adriani2014} Adriani,~O.,~\etal,
\Journal{\ApJ}{791}{93}{2014}
(PAMELA)

\bibitem{Yoon2011} 
Yoon,~H.,~S.,~\etal,
\Journal{\ApJ}{715}{1400--1407}{2011}
(CREAM)

\bibitem{Aguilar2010} Aguilar,~M.,~\etal, 
\Journal{\ApJ}{724}{329--340}{2010}
(\AMS-01)

\bibitem{Jones1978} Jones, F., 
\Journal{\ApJ}{222}{1097}{1978}

\bibitem{Letaw1983} Letaw,~J.~R., \etal, 
\Journal{\ApJ}{51}{271}{1983}

\bibitem{Ahlers2009} Ahlers, M., \etal, 
\Journal{\PRD}{80}{123017}{2009}

\bibitem{TomassettiDonato2012} Tomassetti,~N., \& Donato,~F.,
\Journal{\AeA}{544}{A16}{2012} 

\bibitem{Tomassetti2015tsnr} Tomassetti, N., in preparation;

\bibitem{Malkov2012} Malkov,~M.~A., \etal, 
\Journal{\PRL}{108}{081104}{2012}  

\bibitem{Tomassetti2012} Tomassetti,~N., 
\Journal{\ApJ}{752}{L13}{2012} 

\bibitem{Gleeson1968} Gleeson,~L.~J., \&~Axford,~W.~I 
\Journal{\ApJ}{154}{1011}{1968}

\bibitem{Ahn2009} Ahn, H. S., \etal,  
\Journal{ApJ}{707}{593}{2009} (CREAM)

\bibitem{Muller1991} Muller, D., \etal, 
\Journal{\ApJ}{374}{356--365}{1991} (CRN)

\bibitem{Ave2008} Ave, M., \etal, 
\Journal{ApJ}{678}{262}{2008} (TRACER/03)

\bibitem{Obermeier2011} Obermeier, A., \etal, 
\Journal{\ApJ}{741}{14}{2011} (TRACER/06)

\bibitem{Binns1988} Binns, W.~R., \etal,
\Journal{\ApJ}{324}{1106-1117}{1988}

\bibitem{Seo2014} Seo, E. S., \etal,
\Journal{Ad.Sp.Res.}{53}{1451}{2014} (ISS-CREAM)

\bibitem{Adriani2014Calet} Adriani, O., \etal, 
\Journal{Nucl.Phys.B}{256}{225}{2014} (CALET) 

\bibitem{Moskalenko2003} Moskalenko,~I.~V. \etal, 
\Journal{\ApJ}{586}{1050}{2003}

\bibitem{Benitez2002} Benitez, N., \etal, 
\Journal{\PRL}{88}{081101}{2002}



\end{thebibliography}
\end{document}